\documentclass[epj]{webofc}
\usepackage[utf8]{inputenc}
\usepackage[varg]{txfonts}   % Web of Conferences font
\usepackage{booktabs}
\usepackage{xcolor}
\usepackage{placeins}
\definecolor{darkred}{rgb}{0.4,0.0,0.0}
\definecolor{darkgreen}{rgb}{0.0,0.4,0.0}
\definecolor{darkblue}{rgb}{0.0,0.0,0.4}
\usepackage[bookmarks,linktocpage,colorlinks,
    linkcolor = darkred,
    urlcolor  = darkblue,
    citecolor = darkgreen]{hyperref}
\usepackage{subfigure}
\wocname{EPJ Web of Conferences}
\woctitle{Lattice2017}
\newcommand{\A}{{a}} 
 
\newcommand{\gGF}{g_{\rm GF}^2}
\newcommand{\MSb}{\overline{\textrm{MS}}}
\newcommand{\dd}{\textrm{d}}

\begin{document}
%%%%%%%%%%%%%%%%%%%%%%%%%%%%%%%%%%%%%%%%%%%%%%%%%%%%%%%%%%%%%%%%%%%%%%%%%%%%%
%
\selectlanguage{english}
%----------------------------------------------------------------------------
\title{%
Conformal window in SU(2) with fundamental fermions
}
%----------------------------------------------------------------------------
\author{%
\firstname{Viljami} \lastname{Leino}\inst{1}\fnsep\thanks{Speaker, \email{viljami.leino@helsinki.fi}}
\and
\firstname{Jarno} \lastname{Rantaharju}\inst{2} 
\and
\firstname{Teemu} \lastname{Rantalaiho}\inst{1} 
\and
\firstname{Kari} \lastname{Rummukainen}\inst{1} 
\and
\firstname{Joni} \lastname{Suorsa}\inst{1} 
\and
\firstname{Kimmo} \lastname{Tuominen}\inst{1} 
\and
\firstname{Sara} \lastname{Tähtinen}\inst{1} 
}
%----------------------------------------------------------------------------
\institute{%
University of Helsinki and Helsinki Institute of Physics
\and
Duke University
}
%----------------------------------------------------------------------------
\abstract{%
  We present the updated results of the infrared behavior of the SU(2) model with 6 and 8 fundamental representation fermions. 
  We use the gradient flow method with the Schrödinger functional boundary conditions to measure 
  the running of the coupling in these theories and find fixed points on both. 
  We also measure the mass anomalous dimension from these configurations. 
  %and compare those results to mass anomalous dimension measurements from the mass spectrum simulations.
}
%----------------------------------------------------------------------------
\maketitle
%----------------------------------------------------------------------------
\section{Introduction}\label{intro}
The search for phenomenologically viable models for beyond the standard model scenarios 
has led to a studies of a vacuum phase of $\mathrm{SU}(N_c)$ gauge theories as a function of $N_f$ massless flavors of Dirac fermions.  
At a small $N_f$ these theories are confining and break the chiral symmetry analogously to the QCD. 
On the other hand, at large $N_f^\mathrm{AF}=11 N_c /2$ the theory loses asymptotic freedom and the Gaussian fixed point
is no longer attractive in the ultraviolet. 
Below $N_f^\mathrm{AF}$ the theory develops a non-trivial infrared fixed point (IRFP)
at which only the mass stays a relevant coupling and all the other operators, including gauge coupling, become irrelevant.
While the theory has an IRFP, the theory is conformal i.e. scale invariant and no confinement or chiral symmetry breaking
can occur. When the $N_f$ is lowered, the IRFP moves to the stronger coupling
until at $N_F^C$ the theory becomes chirally broken. 
The interval $N_f^C < N_f < N_f^\mathrm{AF}$ is called the conformal window. 
The lower edge of conformal window, marked by $N^C_f$, is of a special interest as the IRFP typically occurs at large coupling mandating
the use of non-perturbative methods such as lattice simulations. 
Over recent years, the infrared behaviour of multiple different models with varying $N_c$, $N_f$, 
and different fermion representations have been heavily studied on the lattice. 
For recent reviews, see~\cite{Pica:2017gcb,Svetitsky:2017xqk}.
%~\cite{DeGrand:2015zxa,Nogradi:2016qek,Pica:2017gcb,Svetitsky:2017xqk}.

In this paper we focus on the infrared behavior of $\mathrm{SU}(2)$ gauge theories, with varying number of fundamental representation
massless Dirac fermions. With $N_f=2$ the chiral symmetry is broken
and the theory forms a basic template for a dynamical electroweak symmetry breaking~\cite{Hietanen:2014xca}.
Likewise, the lattice studies on $N_f=4$ indicate the theory to be chirally broken~\cite{Karavirta:2011zg}.
Different approximations estimate the lower boundary of conformal window to be 
$N_f^C\sim6-8$~\cite{Sannino:2004qp,Dietrich:2006cm}.
However, both the $N_f=6$ and $N_f=8$ have been controversial, 
with previous studies being inconclusive~\cite{Ohki:2010sr,Karavirta:2011zg,Bursa:2010xn,Hayakawa:2013maa,Appelquist:2013pqa}.
The $N_f=10$ was confirmed to have an IRFP in~\cite{Karavirta:2011zg} and above $N_f=11$ the theory loses its asymptotic freedom.

In this paper we review the results obtained in~\cite{Leino:2017lpc,Leino:2017hgm}. 
%with preliminary results presented 
%in previous conferences~\cite{Rantaharju:2014ila,Suorsa:2015hoh,Leino:2015bfg,Suorsa:2016jsf,Leino:2016njf}.
In these studies we see a clear indication for the existence of IRFP in the $\mathrm{SU}(2)$ gauge theory 
for both $N_f=6$ and $N_f=8$. We also report measurements for the gauge invariant quantities, the mass anomalous dimension 
$\gamma_m^\ast$ and the leading irrelevant exponent $\gamma_g^\ast$ related to the slope of $\beta$-function.
%We do a thorough study of the step scaling function using the gradien flow method~\cite{Luscher:2011bx,Luscher:2010iy} 
%with Schrödinger functional boundary conditions~\cite{Luscher:1991wu,Fritzsch:2013je} to inspect the running of the coupling. 
%----------------------------------------------------------------------------
\section{Lattice setup}\label{sec-1}
We model the SU(2) gauge theory with varying number of massless Dirac fermions
in the fundamental representation using a lattice action:
\begin{equation} 
	S = (1-c_g)S_G(U) + c_g S_G(V) + S_F(V) + c_{SW} \delta S_{SW}(V)\,,
\end{equation}%
where the smeared gauge link $V$ is defined by the hypercubic truncated stout smearing (HEX)~\cite{Capitani:2006ni},
and the smeared, $S_g(V)$, and unsmeared, $S_G(U)$, Wilson gauge actions are mixed together with mixing coefficient
$c_g=0.5$. The clover fermion action, $S_F(V)$, is improved to order $\mathcal{O}(a)$, $a$ being the lattice spacing,
by the use of tree-level Sheikholeslami-Wohlert coefficient $c_{SW}= 1$. 
We use the Schr\"odinger Functional method~\cite{Luscher:1991wu} 
with Dirichlet boundary conditions at the temporal boundaries $x_0=0,L$.
%\begin{equation}
%\begin{aligned}
%U_k(0,{\bf{x}})&=U_k(L,{\bf{x}})=V_k(0,{\bf{x}})=V_k(L,{\bf{x}})=1 \,,\\
%U_\mu(x_0,{\bf{x}}+L\hat{{\bf{k}}})&=U_\mu(x_0,{\bf{x}})\,,\; V_\mu(x_0,{\bf{x}}+L\hat{{\bf{k}}})=V_\mu(x_0,{\bf{x}})\,,\\
% \psi(0,{\bf{x}})&=\psi(L,{\bf{x}})=0\,,\; \psi(x_0,{\bf{x}}+L\hat{{\bf{k}}}) = \psi(x_0,{\bf{x}}) \,
%\end{aligned}
%\end{equation}%

These boundary conditions allow us to run our simulations at zero quark mass. 
Thus we define $\kappa_c(\beta_L)$ as the value at which the PCAC mass~\cite{Luscher:1996vw} vanishes.
We tune the value of $\kappa_c(\beta_L)$ on lattices of size $L=24$ for each $\beta_L$ in use by
doing multiple measurements of the PCAC mass with different values of the hopping parameter $\kappa$ and interpolating to zero mass.
We attain an accuracy of $10^{-5}$. 

To run our simulations we utilize the hybrid Monte Carlo algorithm 
with 2nd order Omelyan integrator~\cite{Omelyan:2003:SAI}.
The step length is tuned to achieve an acceptance ratio above 85\%. 
We follow the evolution of topological charge during the simulations and make sure 
to only do analysis on trajectories free of topological freezing.
In general we generate $(5-100)\cdot 10^3$ trajectories for each combination of $\beta_L$ and $L$ used.
The comprehensive algorithmic details are described in~\cite{Rantaharju:2015yva,Leino:2017lpc}.

\section{Theory}\label{sec-2}
\subsection{Running of the coupling}\label{subsec-2.1}
We measure the coupling using the Yang-Mills gradient flow~\cite{Luscher:2009eq,Luscher:2010iy}:
\begin{equation}
  \partial_t B_{\mu} = D_{\nu} G_{\nu\mu} \,,
  \label{eq:gradflow}
\end{equation}%
where the initial condition is $B_{\mu}(x;t=0) = A_\mu(x)$, $D_\mu=\partial_\mu+[B_\mu,\,\cdot\,]$
is the covariant derivative of the field strength tensor $G_{\mu\nu}(x;t)$
and $t$ is the fictitious flow time.
This smearing transformation drives the gauge field towards the minima of Yang-Mills action and 
continuously removes the UV divergences by smoothing the gauge field. 
The gradient flow automatically renormalizes gauge invariant objects,
allowing us to measure the coupling at a scale $\mu=1/\sqrt{8t}$ as:
\begin{align}
	\label{eq:g2gf}
	\gGF(\mu) &= \mathcal{N}^{-1}t^2 \langle E(t+\tau_0 a^2) \rangle\vert_{x_0=L/2\,,\,t=1/8\mu^2}\,,
\end{align}%
where the $\tau_0$ is a tunable shift parameter introduced in~\cite{Cheng:2014jba}
to reduce the $\mathcal{O}(a^2)$ discretization effects. 
We use the normalization factor $\mathcal{N}$ defined in~\cite{Fritzsch:2013je}
to match the gradient flow coupling $\gGF$ to the $\MSb$ coupling in the tree level.
As the translation symmetry is broken by the chosen boundary conditions,
the coupling is only measured along the central time slice. 

The finite size and cutoff effects in gradient flow coupling depend on
the discretizations chosen for the action used for the simulations, the action minimized by the flow,
definition of the energy density $E(t)$ and the chosen boundary conditions.
Therefore, the scale $\mu$ is commonly limited to a regime where both the cutoff effect and statistical variance,
that tends to increases with larger flow times, are reasonably small. 
This is achieved by choosing a dimensionless parameter $c_t=\sqrt{8t}/L=L/\mu$~\cite{Fritzsch:2013je,Fodor:2012td},
which defines the renormalization scheme.

The evolution of the coupling is then quantified by the step scaling function:
\begin{equation} \label{eq:lat_step_raw}
    \Sigma(u,L/\A,s) = \left . \gGF(g_0,sL/\A) \right|_{\gGF(g_0,L/\A)=u}\,,
\end{equation}%
which describes the change in the measured coupling $\gGF$ caused by the increase in the lattice size
from $L$ to $sL$. 
As we use $\mathcal{O}(a)$ improved actions, 
we expect the lowest order discretization effects to be of order $\mathcal{O}(a^2)$
and extrapolate the continuum step scaling function with a fit:
\begin{align} \label{eq:lat_step_cont}
	\Sigma(u,L/\A)  &= \sigma(u) + c(u) (\A/L)^2\,.
\end{align}%

In the proximity of an IRFP the continuum step scaling function $\sigma$ can be related to the $\beta$-function as:
\begin{align}
\beta(g) = -\mu \frac{dg^2}{d\mu} = \gamma_g^\ast(g^2-g_\ast^2) 
\approx 
\frac{g}{2\ln(s)} \left ( 1 - \frac{\sigma(g^2,s)}{g^2} \right)\,.
\label{eq:betaas}
\end{align}%
This allows us to measure the slope of the $\beta$-function depicting the leading irrelevant exponent $\gamma_g^\ast$.
However, the existence of an IRFP can introduce scaling violations into the typical Symanzik type improvements,
bringing into question the validity of the continuum limit~\eqref{eq:lat_step_cont} at the IRFP.
To check the reliability of the $\gamma_g^\ast$ measurement from the slope, we compare it to the alternative 
finite size scaling method devised in~\cite{Appelquist:2009ty,DeGrand:2009mt,Lin:2015zpa,Hasenfratz:2016dou} and fit:
\begin{equation}
\gGF(\beta,L) - g_\ast^2 = \left[ \gGF(\beta,L_\mathrm{ref})-g_\ast^2\right]\left(\frac{L_\mathrm{ref}}{L}\right)^{\gamma_g^\ast}\,.
\label{eq:ramosgamma}
\end{equation}
Here $L_\mathrm{ref}<L$ is in principle arbitrary reference lattice size, and $\gamma_g^\ast$ should be independent of it
if $\gGF$ is sufficiently close to the IRFP.

\subsection{Mass anomalous dimension}\label{subsec-2.2}
Our choice of boundary conditions allows us to also measure the mass anomalous dimension $\gamma_m^\ast$ using 
two different methods, the mass step scaling method and the spectral density method.
In the step scaling method we measure the running of 
the pseudoscalar density renormalization constant~\cite{Capitani:1998mq,DellaMorte:2005kg} as:
\begin{align}
Z_P(g_0,L) = \frac{\sqrt{N_c f_1} }{f_P(L/2)}\,,
\label{Zp}
\end{align}%
and define the $\gamma_m^\ast$ from the step scaling function:
\begin{align}
  \gamma_m^\ast(u) = -\frac{\log \sigma_P(u,s)}{\log s }\,,\quad\mathrm{where}\quad 
  \sigma_P(u,s)=\lim_{a\rightarrow\infty} \left. \frac {Z_P(g_0,sL/a)}{Z_P(g_0,L/a)} \right |_{g_{\rm{GF}}^2(g_0,L/a)=u}
\label{eq:gammastar_zp}
\end{align}%

The second way to measure $\gamma_m^\ast$ is based on the fact that the scaling of the spectral density
of massless Dirac operator is also determined by it.
Using the recently introduced stochastic methods 
it is possible to extract the mode number of Dirac operator~\cite{Giusti:2008vb,Patella:2011jr}
from its eigenvalue density $\rho(\lambda)$.
It is known that in the vicinity of an IRFP the mode number follows a scaling behavior:
\begin{equation}\label{modenumber1}
\nu(\Lambda) \equiv 2\int_0^{\sqrt{\Lambda^2 - m^2}} \rho(\lambda)\dd \lambda
\simeq \nu_0(m) +  C\left[\Lambda^2 - m^2\right]^{2/(1+\gamma_\ast)}
%\propto 
\simeq \Lambda^{4/(1+\gamma_m^\ast)}\,,
\end{equation}%
where we have assumed $\nu_0(m)$ and $m^2$ to be small due to vanishing PCAC quark masses.

\section{Measurements}\label{sec-3}
We study the $\mathrm{SU}(2)$ gauge theory with $N_f=6$ and $N_f=8$. 
For the $N_f=6$ we use the lattice sizes of $L=8,10,12,16,18,20,24,30$ and choose the step-size in Eqs~\eqref{eq:lat_step_raw}
and~\eqref{eq:gammastar_zp} to be $s=3/2$. On the other hand, the $N_f=8$ analysis is done
with $L=6,8,10,12,16,20,24,32$ and $s=2$. 

\subsection{Running of the coupling}\label{subsec-3.1}
\begin{figure}[thb] % no figure before 1st section
  \centering
  \includegraphics[width=0.49\linewidth,clip]{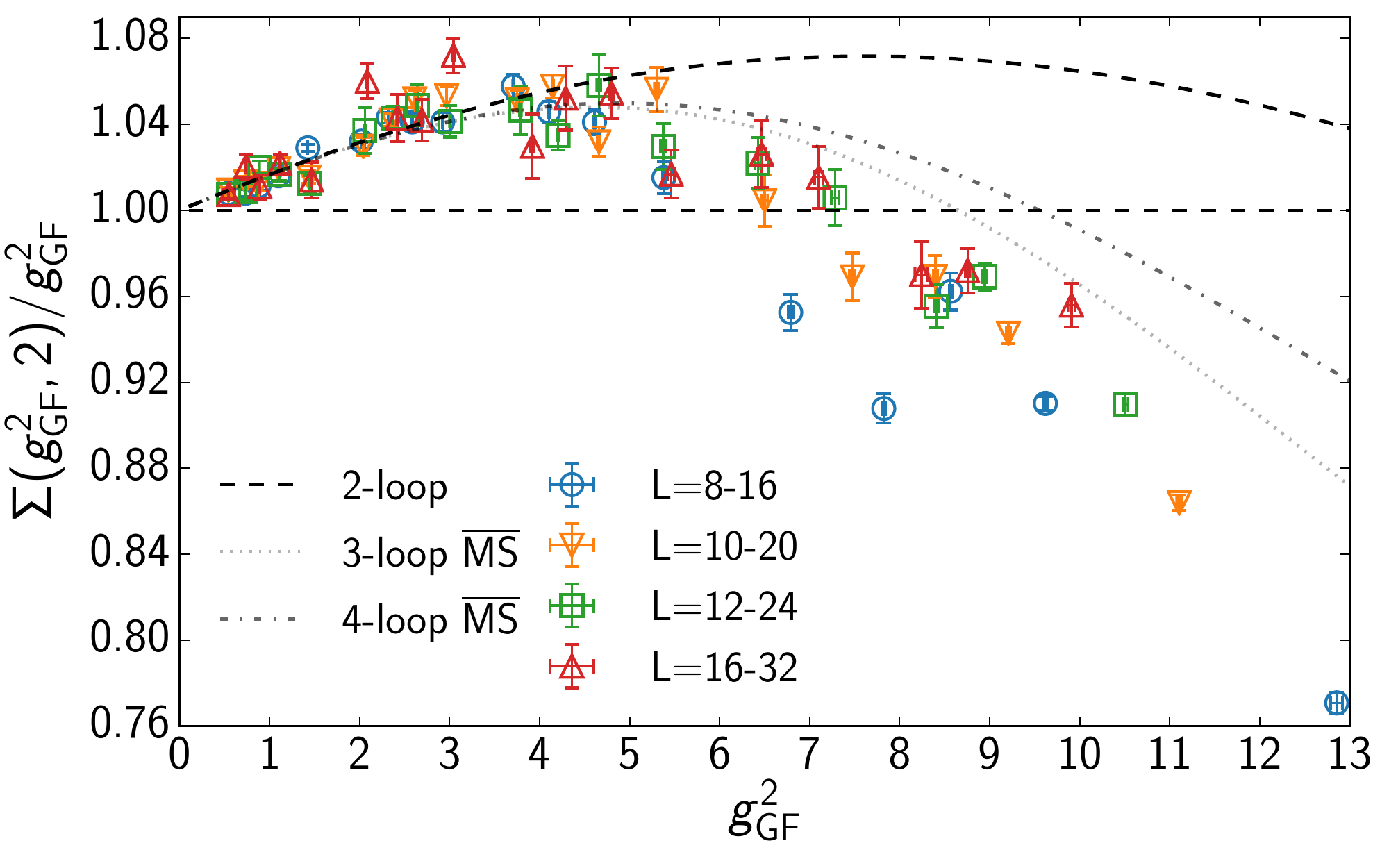}
  \includegraphics[width=0.49\linewidth,clip]{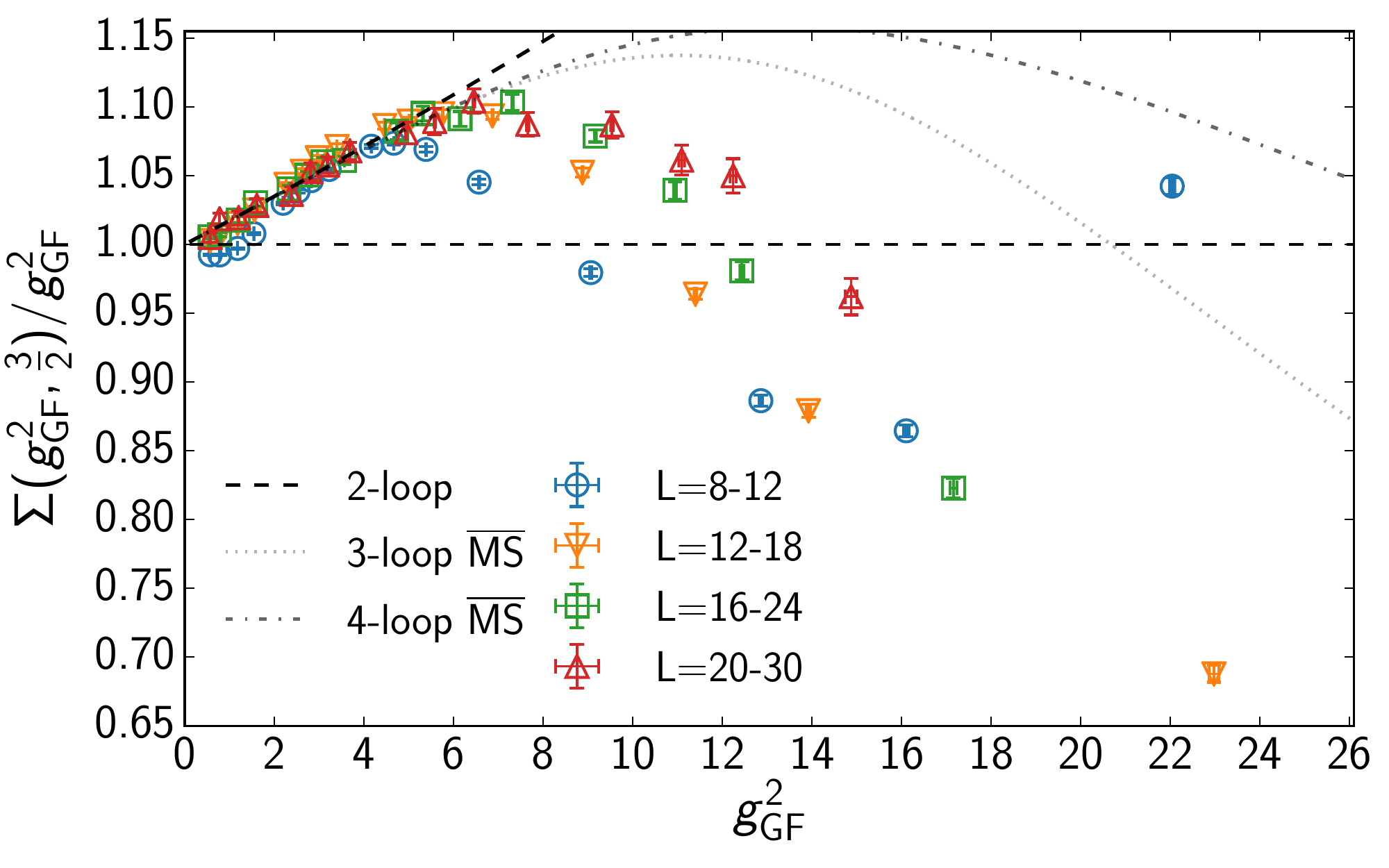}
  \caption{The lattice step scaling function $\Sigma(g_0^2,s)$ with $\tau_0$-correction for {\em Left:} $N_f=8$ at $c_t=0.4$ 
  and {\em Right:} $N_f=6$ at $c_t=0.3$.}
  \label{fig-2}% Give a unique label
\end{figure}

We measure the coupling using the gradient flow~\eqref{eq:gradflow} method,
for which we have chosen the L\"uscher-Weisz action for the flow discretization and
the clover discretization for the energy $E(t)$.
The scale is set by the choice $c_t=0.3$ for $N_f=6$ and $c_t=0.4$ for $N_f=8$. 
We present the lattice step scaling function~\eqref{eq:lat_step_raw} for both $N_f=6$ and $N_f=8$ in Figure~\ref{fig-2}
and observe that at small couplings the results agree with the perturbative curves until
they start to curve towards the IRFP at large couplings.
On large couplings, the smallest lattice pair on both $N_f=6$ and $N_f=8$ seems to have slightly different behavior
than the rest of the lattice pairs likely caused by finite size effects.
Because of this lattices smaller than $L=10$ are excluded from all further analysis.
\begin{figure}[thb] % no figure before 1st section
  \centering
  \includegraphics[width=0.49\linewidth,clip]{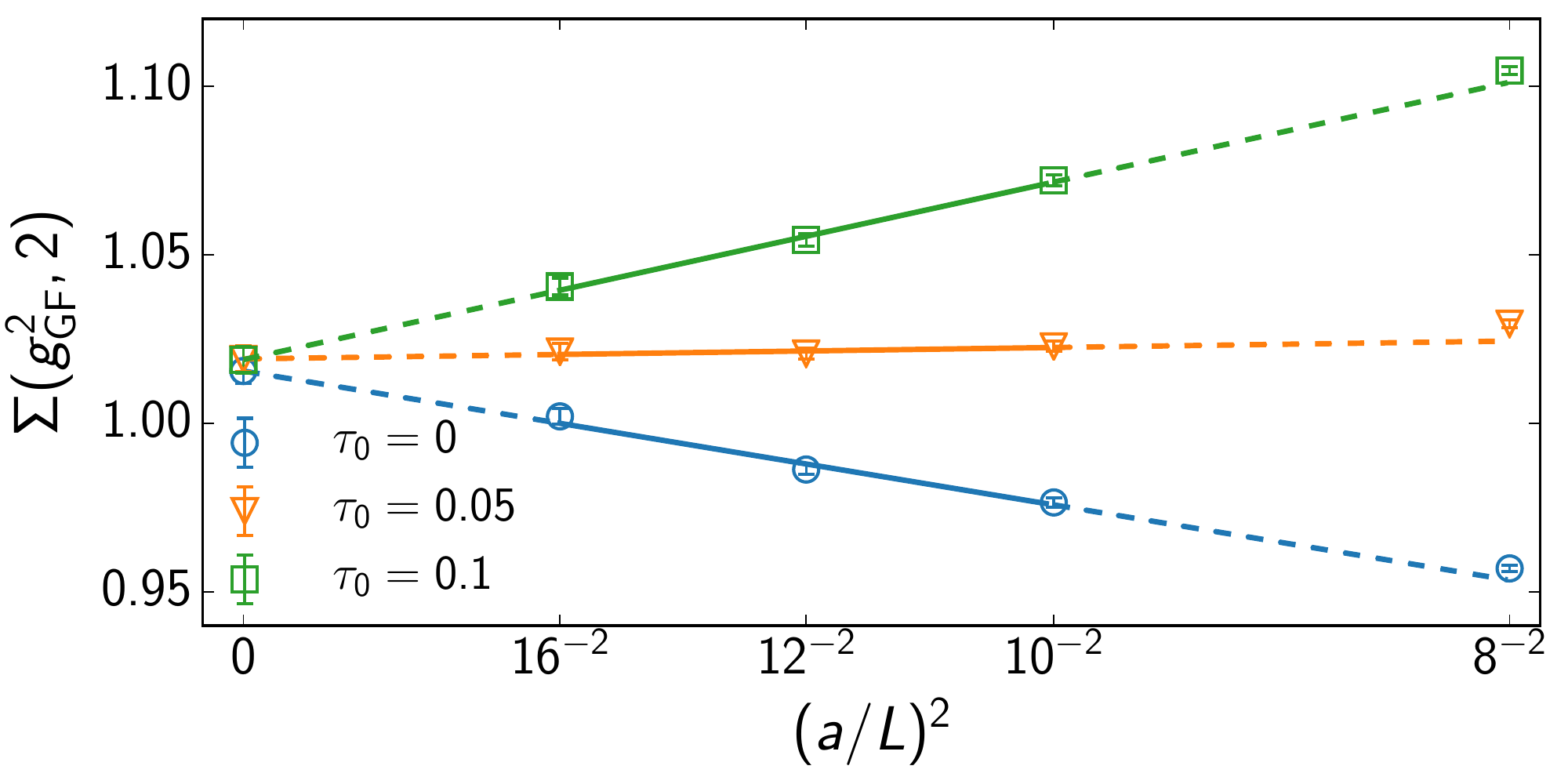}
  \includegraphics[width=0.49\linewidth,clip]{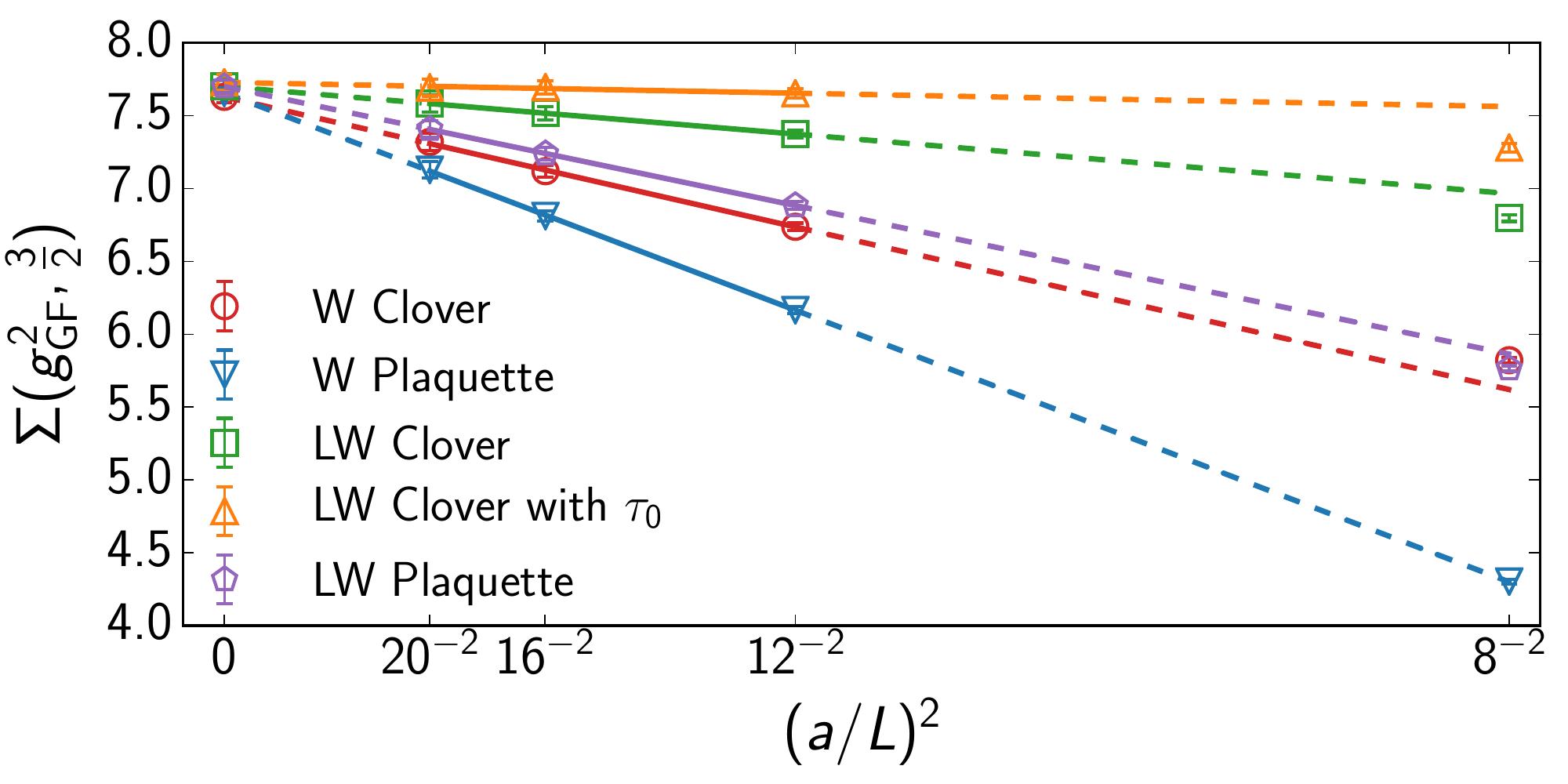}
  \caption{{\em Left:} The effect of $\tau_0$-correction demonstrated using the continuum limit of the step scaling function in $N_f=8$ 
  theory at $\gGF=1$.
  {\em Right:} The effect of different discretizations to the continuum limit demonstrated
  using the continuum step scaling function in $N_f=6$ theory at $\gGF=7$.
  In the legend LW refers to L\"uscher-Weisz flow, W refers to Wilson flow and plaquette and clover are the different discretizations 
  for $E(t)$.
  }
  \label{fig-3}% Give a unique label
\end{figure}%
\begin{figure}[thb] % no figure before 1st section
  \centering
  \includegraphics[width=0.49\linewidth,clip]{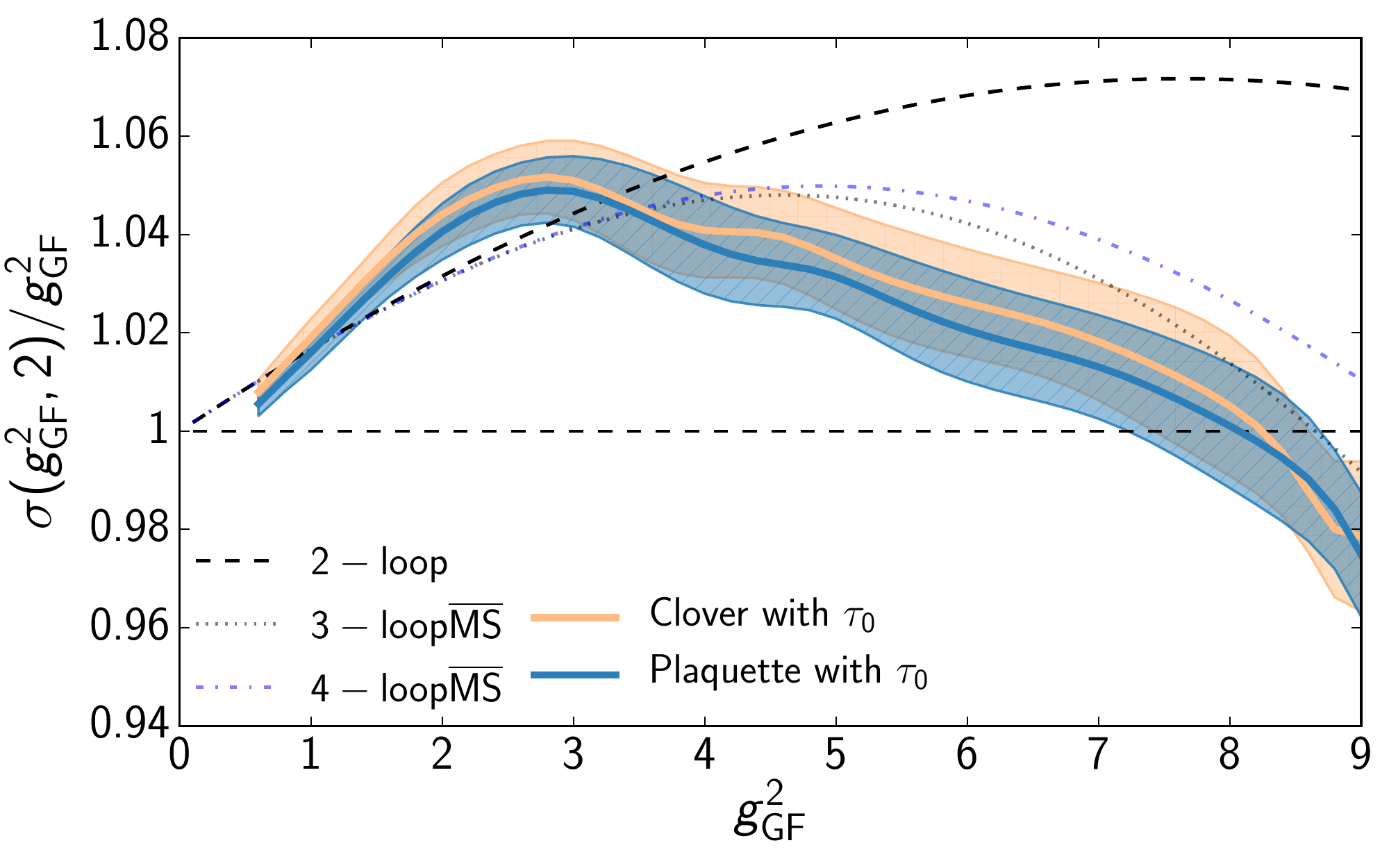}
  \includegraphics[width=0.49\linewidth,clip]{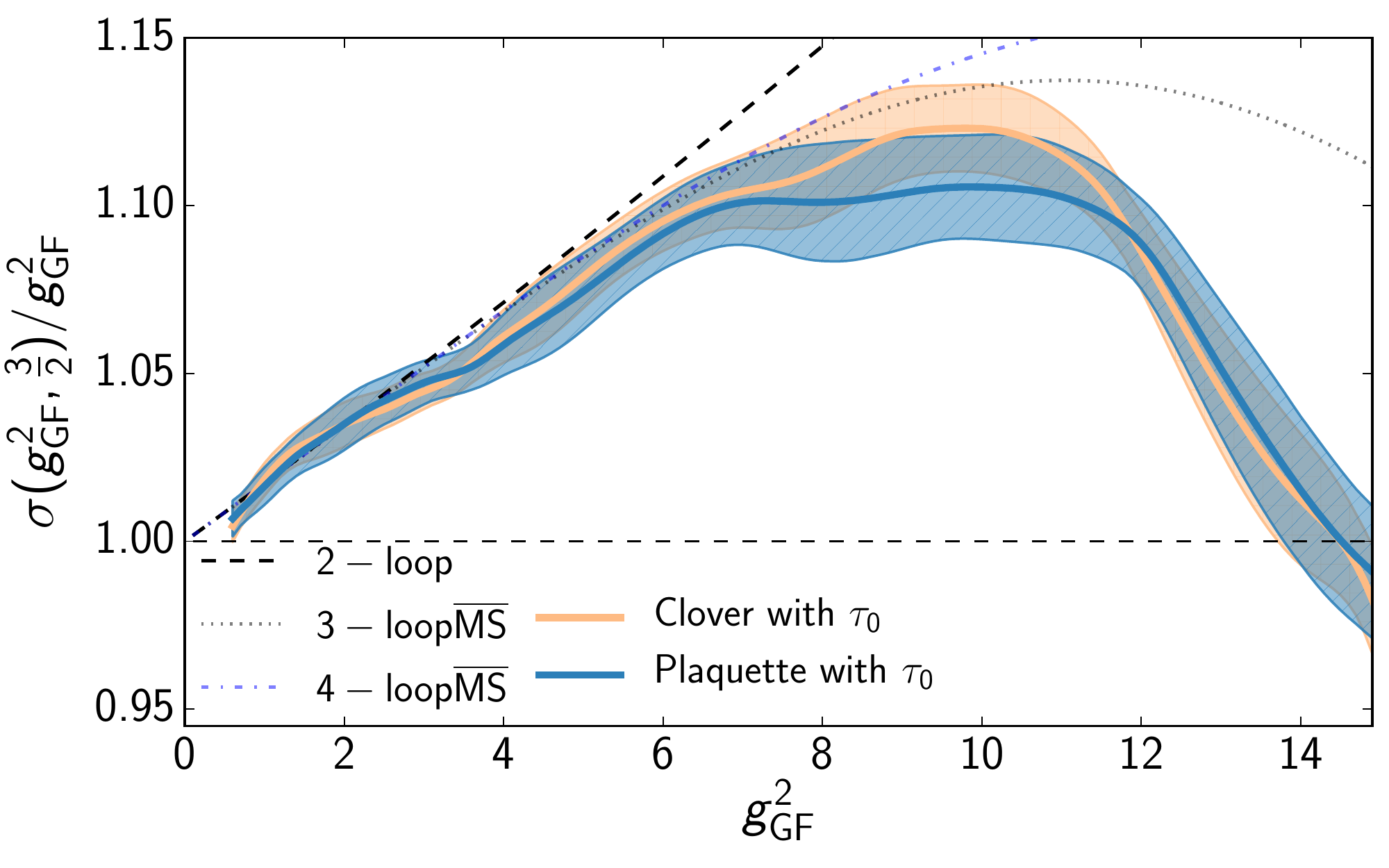}
  \caption{The continuum limit of the step scaling function $\sigma(\gGF,s)$ with different discretizations for 
  the $E(t)$ for {\em Left:} $N_f=8$ at $c_t=0.4$ and {\em Right:} $N_f=6$ at $c_t=0.3$.
  The errors include statistical errors and systematics rising from choice of interpolation function.}
  \label{fig-5}% Give a unique label
\end{figure}%
\begin{figure}[thb] % no figure before 1st section
  \centering
  \includegraphics[width=0.49\linewidth,clip]{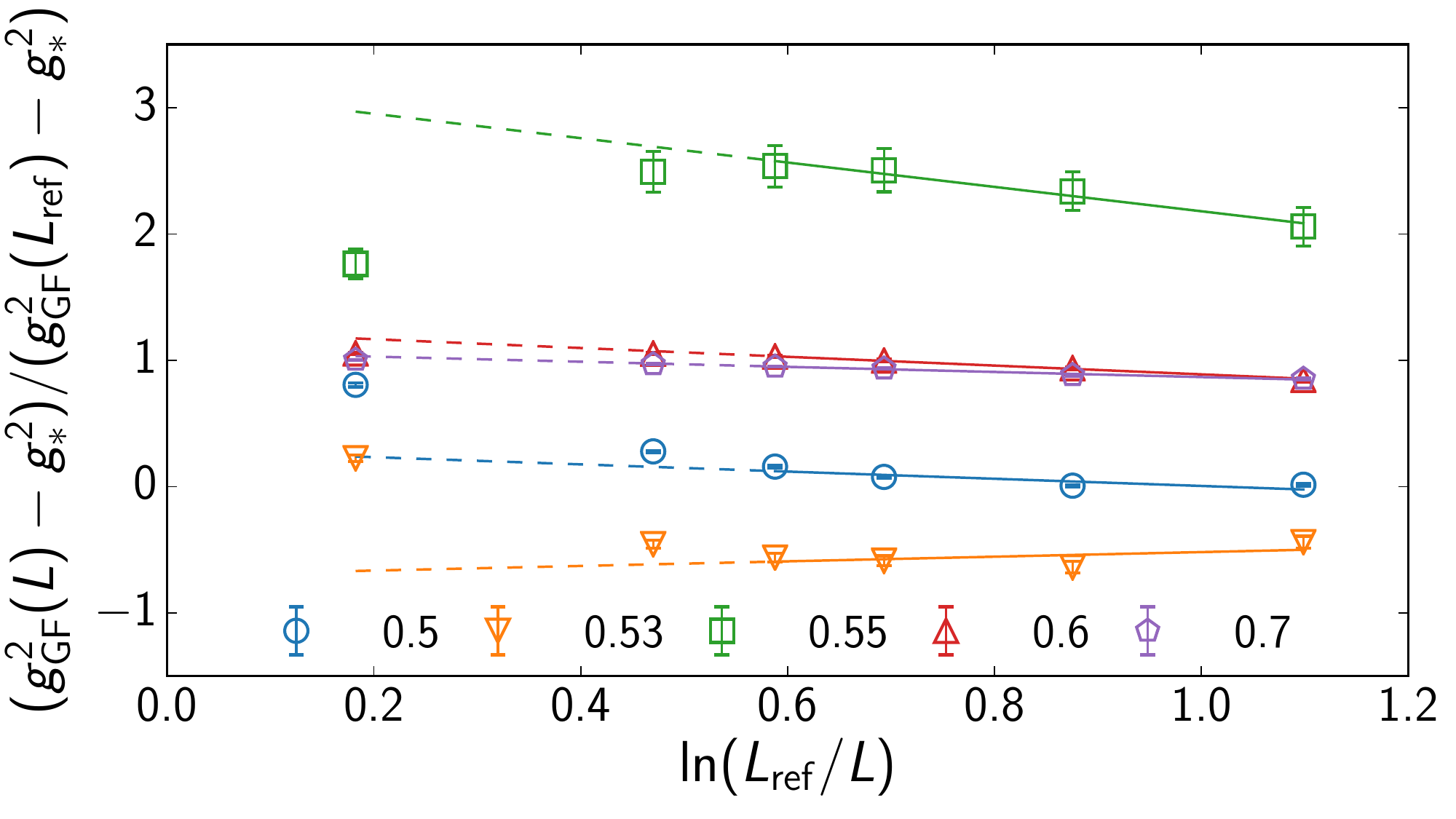}
  \includegraphics[width=0.49\linewidth,clip]{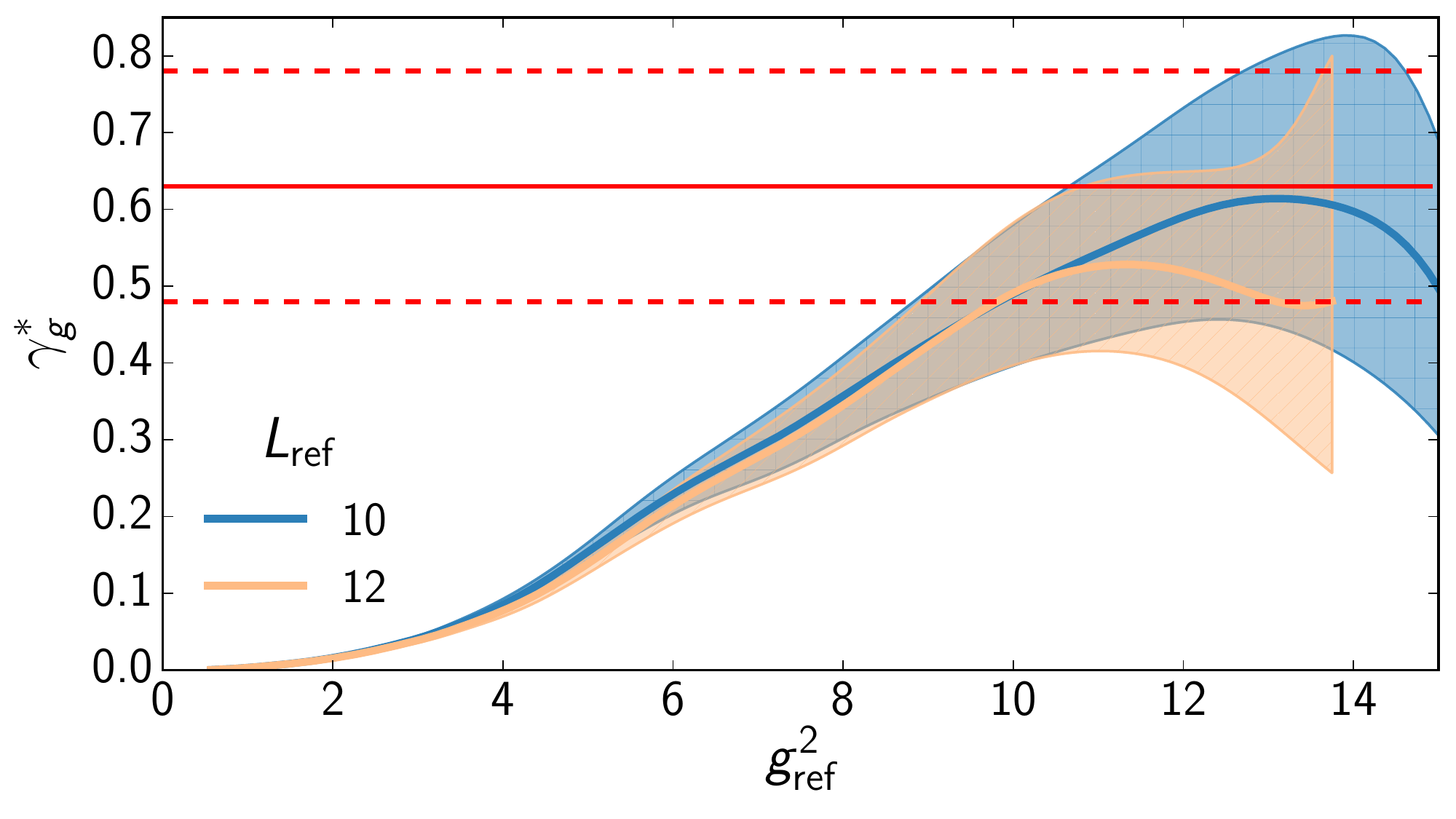}
  \caption{The alternative $\gamma_g^\ast$ fit~\eqref{eq:ramosgamma},  {\em Left:} For 5 smallest $\beta$'s.
  {\em Right:} For all interpolated values of $g_\mathrm{ref}^2$.
  }
  \label{fig-6}% Give a unique label
\end{figure}%

We tune the $\tau_0$ correction by minimizing the $\mathcal{O}(a^2)$ effects as shown 
on the left side of Figure~\ref{fig-3}.
In general, larger couplings $\gGF$ require larger corrections, but since the $\tau_0$ is assumed to be a small correction we have
chosen a logarithmic form for the $\tau_0$ in order to regulate the strong coupling behavior:
\begin{equation}
\tau_0 = 0.025\log(1+2\gGF)\;,\mathrm{for}\,N_f=6\quad\mathrm{and}\quad\tau_0 = 0.06\log(1+\gGF)\;,\mathrm{for}\,N_f=8
\label{eq:taufunc}
\end{equation}%
On the right hand side of the Figure~\ref{fig-3} we examine the effect of different discretizations to the continuum limit 
alongside the $\tau_0$-correction. Clearly the chosen set of parameters have the smallest discretization effects.
Comprehensive analysis for different parameter choices has been done in~\cite{Leino:2017lpc,Leino:2017hgm}.

We then interpolate the raw couplings, with respect to the bare coupling $g_0^2=4/\beta$, using a 9th degree polynomial 
for the $N_f=6$ and using a rational ansatz with 7th degree polynomial in the numerator 
and 1st degree polynomial in the denominator for the $N_f=8$.
The continuum step scaling function~\eqref{eq:lat_step_cont} is presented in Figure~\ref{fig-5},
where we also ensure the consistency of the continuum limit by checking that the two different discretizations of $E(t)$
have the same continuum limit. Both of the theories are found to exhibit an IRFP at:
$g_\ast^2 = 14.5(3)_{-1.38}^{+0.41}$ for the $N_f=6$ and $g_\ast^2 =8.24(59)_{-1.64}^{+0.97}$ for the $N_f=8$.
Here, the first set of errors is statistical and the second set indicates the systematical uncertainty rising
from varying different discretizations.

From the continuum step scaling function we can also measure the slope of beta function~\eqref{eq:betaas}.
In the $N_f=8$ no proper measurement can be done due to large errors, 
but in $N_f=6$ we can measure $\gamma_g^\ast=0.63(15)_{-0.27}^{+0.28}$ which is consistent with the recent scheme independent estimate
$\gamma_g^\ast=0.6515$~\cite{Ryttov:2017toz}.
For the alternative method~\eqref{eq:ramosgamma} we observe high discretization effects, shown on the left of Figure~\ref{fig-6},
forcing us to only use $L>16$ for the main fit, $L_\mathrm{ref}$ varying between 10 and 12. 
From the right hand side of the Figure~\ref{fig-6}, we can see this method gives comparable values at the IRFP,
although the signal gets weak closer to the IRFP as indicated by the downwards divergence in the figure.

\subsection{Mass anomalous dimension}\label{subsec-3.2}
By interpolating pseudoscalar density renormalization constant $Z_p$~\eqref{Zp} with a 8th order polynomial for the $N_f=6$ and
6th order polynomial for the $N_f=8$, we get the continuum limit of the mass step scaling function~\eqref{eq:gammastar_zp},
shown in Figure~{fig-7}. From the figure we observe that the mass step scaling method becomes unstable at a large couplings
and no IRFP value for the $\gamma_m^\ast$ can be quoted.
\begin{figure}[thb] % no figure before 1st section
  \centering
  \includegraphics[width=0.49\linewidth,clip]{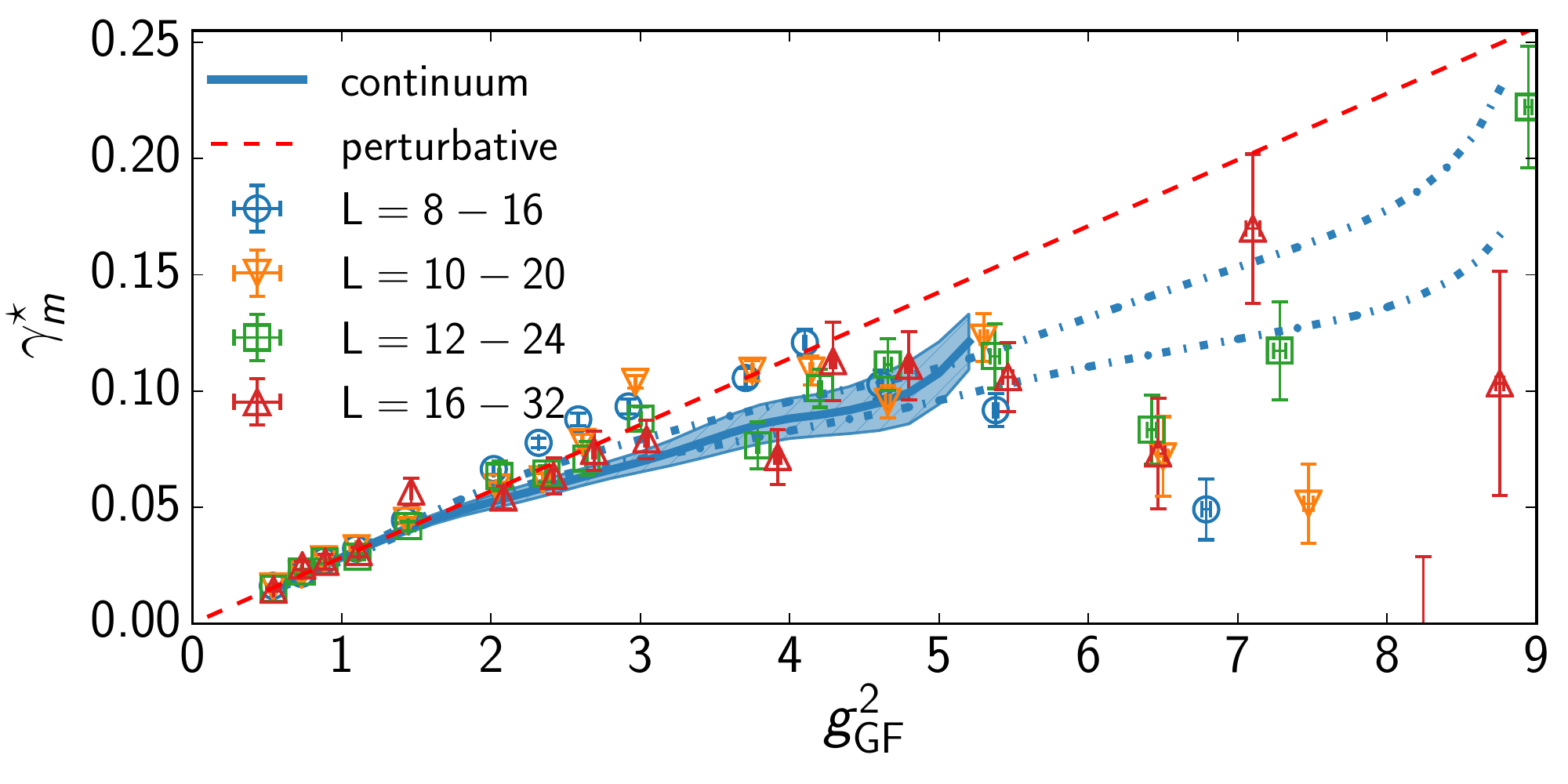} %nf8gammast.pdf
  \includegraphics[width=0.49\linewidth,clip]{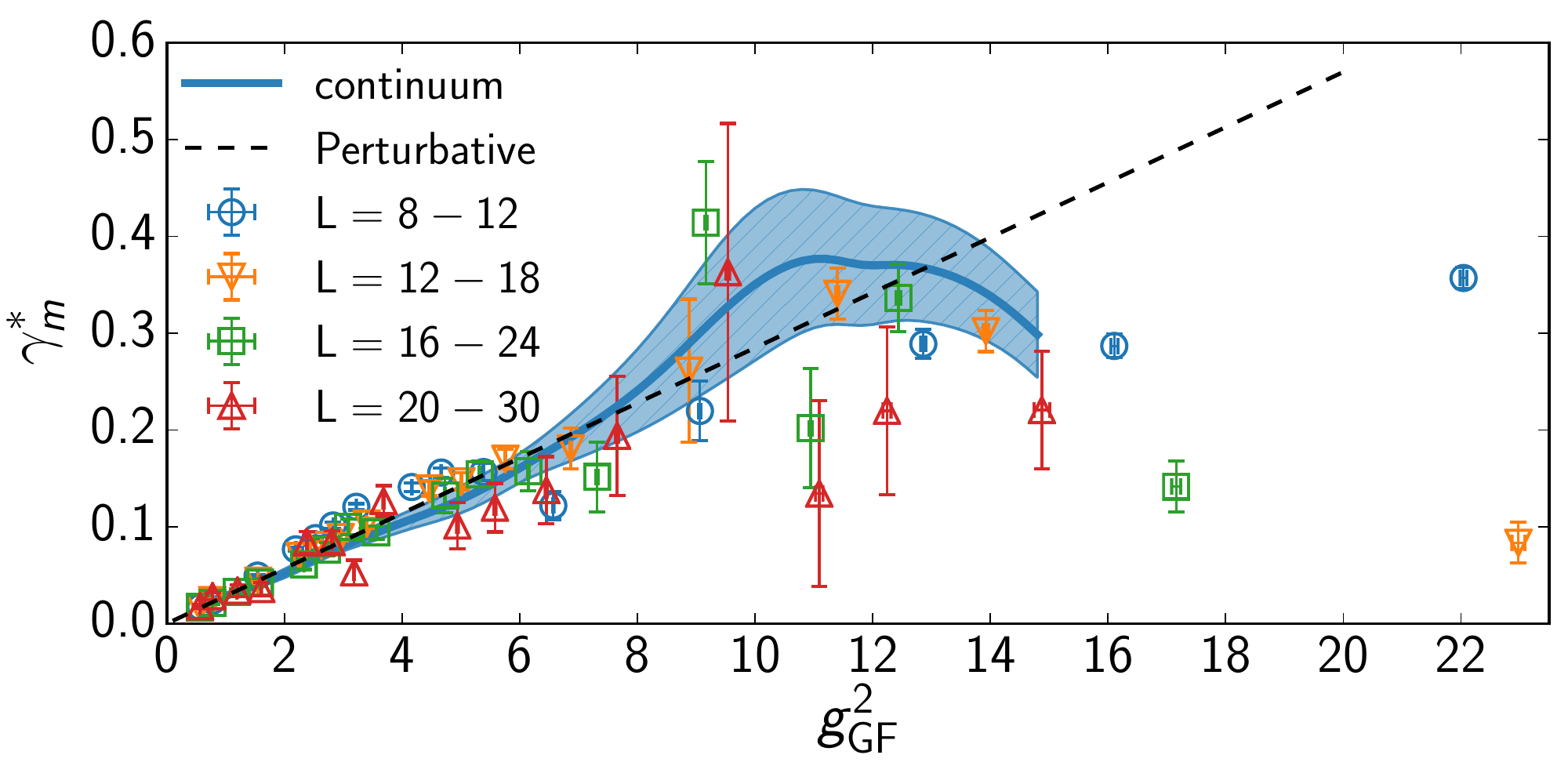} 
  \caption{The mass anomalous dimension $\gamma_m^\ast$ measurement together with its continuum limit~\eqref{eq:gammastar_zp}
  for: {\em Left:} $N_f=8$ and {\em Right:} $N_f=6$. In the region with dashed lines, the fit is unacceptably bad.
  }
  \label{fig-7}% Give a unique label
\end{figure}%
\begin{figure}[thb] % no figure before 1st section
  \centering
  \includegraphics[width=0.49\linewidth,clip]{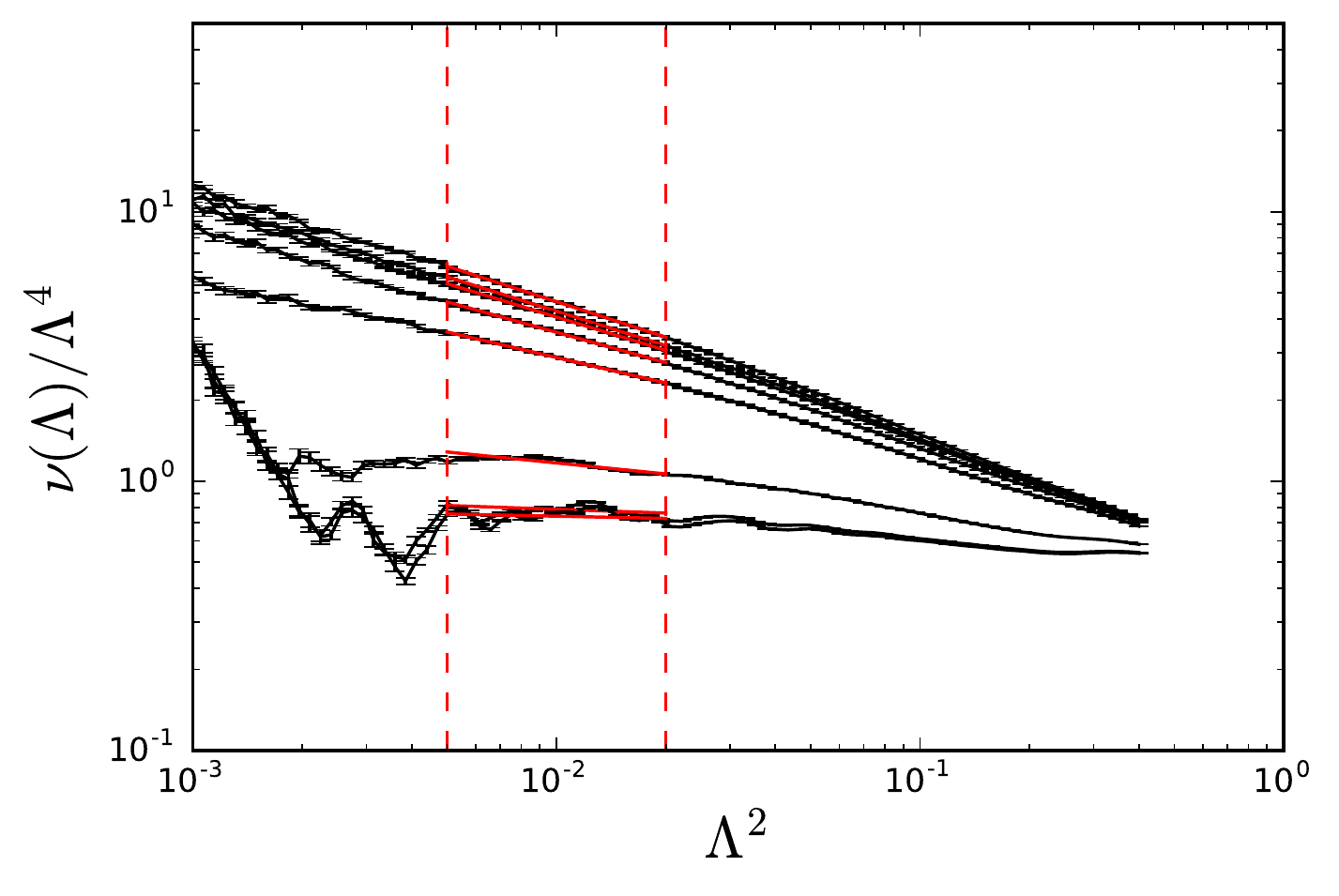}
  \includegraphics[width=0.49\linewidth,clip]{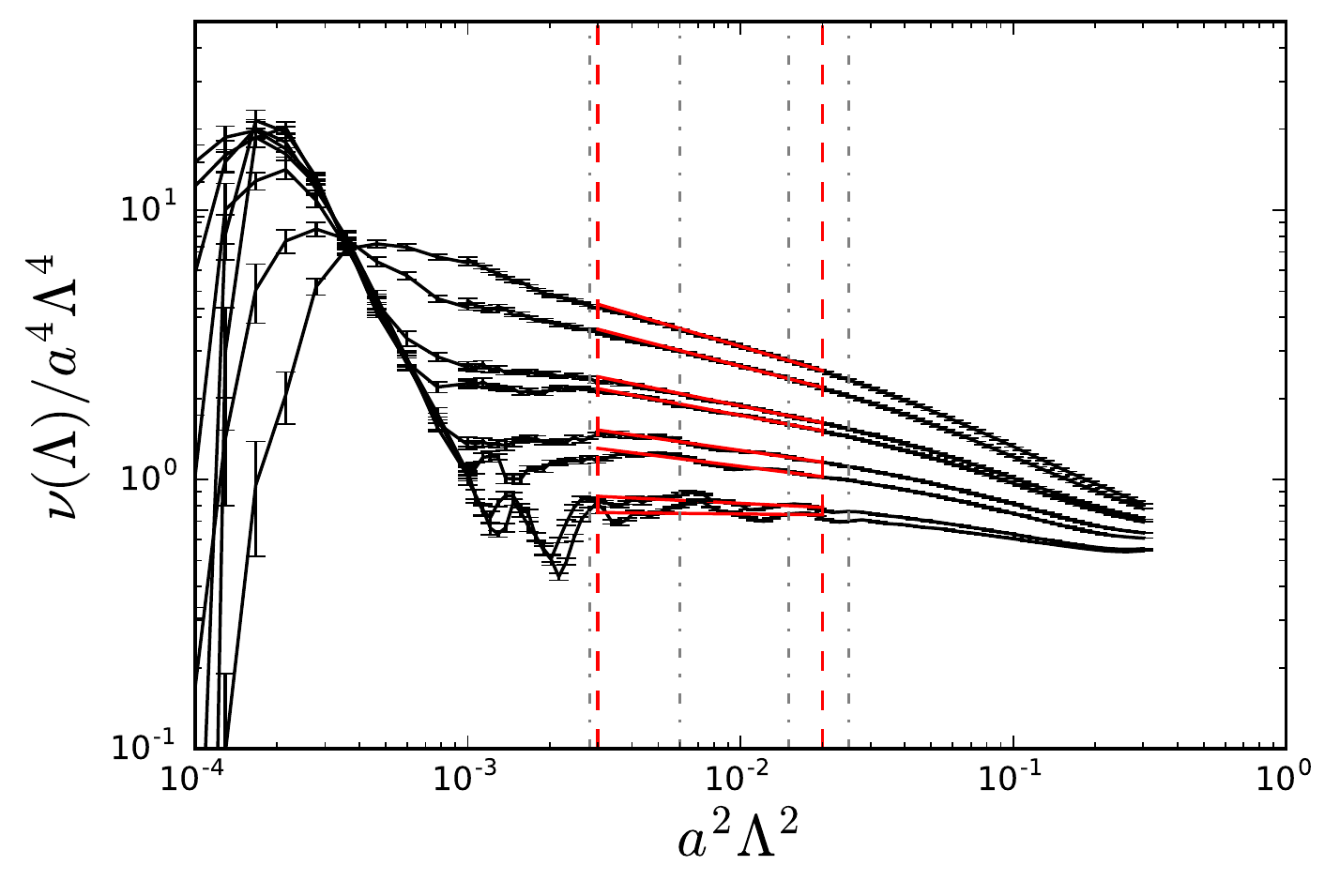} 
  \caption{The mode number divided by $a^4\Lambda^4$ as a function of $a^2\Lambda^2$
  for: {\em Left:} $N_f=8$ and {\em Right:} $N_f=6$.
  The dashed lines indicate the chosen fit range.
  }
  \label{fig-10}% Give a unique label
\end{figure}%
\begin{figure}[thb] % no figure before 1st section
  \centering
  \includegraphics[width=0.49\linewidth,clip]{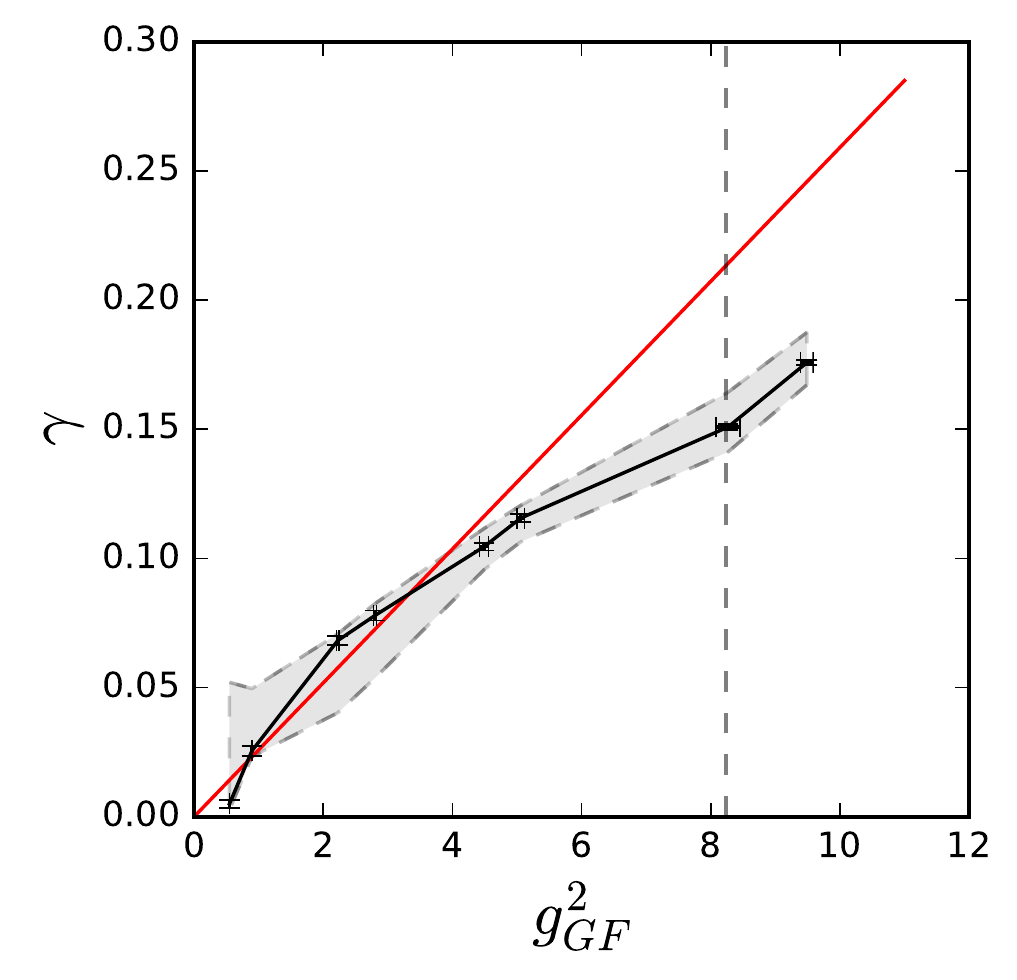}
  \includegraphics[width=0.5\linewidth,clip]{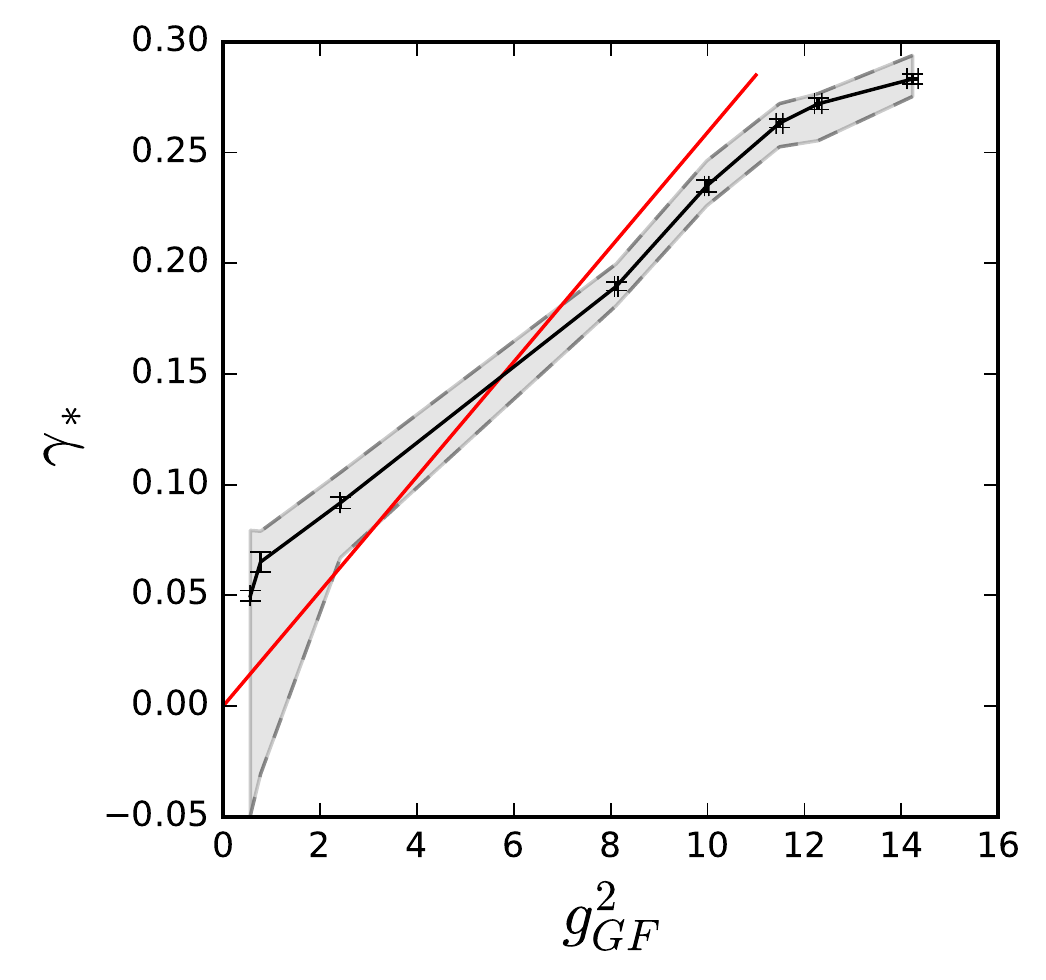} 
  \caption{The mass anomalous dimension $\gamma_m^\ast$ using the spectral density method~\eqref{modenumber1} for
  {\em Left:} $N_f=8$ and {\em Right:} $N_f=6$. The shaded bands illustrates the uncertainty from varying the fit range by 50\%.
  }
  \label{fig-8}% Give a unique label
\end{figure}%

To make a more definitive measurement of the $\gamma_m^\ast$ we turn our focus to the spectral density method.
Taking 10-20 well separated configurations from the step scaling study for each $\beta_L$
at the $L=24$ for $N_f=6$ and $L=32$ for $N_f=8$, 
we calculate the mode number for 90 values of $\Lambda^2$ ranging between $10^{-3}-0.3$.
We measure the $\gamma_m^\ast$ from the mode number by doing the scaling fit~\eqref{modenumber1},
shown in Figure~\ref{fig-10}.
The right scaling is observed at the strong coupling, however, at small coupling the low eigenvalues
appear at discrete energies, which can be seen as bumps in the figure.
The right fit range is determined by cross referencing both perturbation theory and step scaling method at low couplings.

In Figure~\ref{fig-8} we present the final measurement of $\gamma_m^\ast$,
where the shaded bands show the errors arising from varying the fit range in Figure~\ref{fig-10}.
As the bumps at the low coupling make the fit harder, the largest uncertainty from this method 
arises from the low couplings. On the other hand, on large couplings this method turns out to be extremely reliable.
Therefore these results complement the results from the step scaling method perfectly and we can
measure: $\gamma_m^\ast = 0.283(2)^{+0.01}_{-0.01}$ for $N_f=6$ and $\gamma_m^\ast = 0.15(2)$ for $N_f=8$ 
at the IRFP.

\section{Conclusions} 
We have studied the running coupling in the $\mathrm{SU}(2)$ lattice gauge theory 
with 6  and 8 fermions in the fundamental representation and
showed that both theories develop an infrared fixed point at 
$g_\ast^2 = 14.5(3)_{-1.38}^{+0.41}$ and $g_\ast^2 =8.24(59)_{-1.64}^{+0.97}$ for
for $N_f=6$ and $N_f=8$ respectively.
The existence of an IRFP is compatible with our study of the mass spectrum in $\mathrm{SU}(2)$ theories~\cite{Amato:2015dqp}.
We have measured the scheme independent quantities 
$\gamma_m^\ast =0.283(2)^{+0.01}_{-0.01}$ for $N_f=6$ and $\gamma_\ast = 0.15(2)$ for $N_f=8$,
and $\gamma_g^\ast=0.63(15)_{-0.27}^{+0.28}$ for $N_f=6$.

\FloatBarrier
\section{Acknowledgments}
This work is supported by the Academy of Finland grants 267286 and 267842.
V.L. and J.S. acknowledge the funding by Jenny and Antti Wihuri foundation and
T.R. and S.T. are funded by the Magnus Ehrnrooth foundation.
The simulations were performed at the Finnish IT Center for Science (CSC) in Espoo, Finland,
on the Fermi supercomputer at Cineca in Bologna, Italy, and on the K computer at Riken AICS in Kobe, Japan.

%
%\begin{figure}[thb]
%  \centering
%  \sidecaption
%  \includegraphics[width=1.5cm,clip]{logo-lattice2017}
%  \caption{Please write your figure caption here}
%  \label{fig-2}% Give a unique label
%\end{figure}
%
%
%\begin{table}[thb]
%  \small
%  \centering
%  \caption{Please write your table caption here}
%  \label{tab-1}% Give a unique label
%  \begin{tabular}{lll}\toprule
%  first  & second & third  \\\midrule
%  number & number & number \\
%  number & number & number \\\bottomrule
%  \end{tabular}
%\end{table}

% BibTeX or Biber users please use (the style is already called in the class, ensure that the "woc.bst" style is in your local directory)
% \bibliography{name or your bibliography database}
%
%\emph{The figures and tables must appear before the references.}

%----------------------------------------------------------------------------
%\clearpage
%\FloatBarrier
\bibliography{lattice2017}

\begin{thebibliography}{32}

\bibitem{Pica:2017gcb}
C.~Pica, PoS \textbf{LATTICE2016}, 015 (2016), \texttt{1701.07782}

\bibitem{Svetitsky:2017xqk}
B.~Svetitsky, \emph{{Looking behind the Standard Model with lattice gauge
  theory}} (2017), in \emph{Proceedings,
  \href{http://inspirehep.net/record/1425631}{35th International Symposium on
  Lattice Field Theory (Lattice2017)}: Granada, Spain}, to appear in EPJ Web
  Conf., \texttt{1708.04840}

\bibitem{Hietanen:2014xca}
A.~Hietanen, R.~Lewis, C.~Pica, F.~Sannino, JHEP \textbf{07}, 116 (2014),
  \texttt{1404.2794}

\bibitem{Karavirta:2011zg}
T.~Karavirta, J.~Rantaharju, K.~Rummukainen, K.~Tuominen, JHEP \textbf{05}, 003
  (2012), \texttt{1111.4104}

\bibitem{Sannino:2004qp}
F.~Sannino, K.~Tuominen, Phys. Rev. \textbf{D71}, 051901 (2005),
  \texttt{hep-ph/0405209}

\bibitem{Dietrich:2006cm}
D.D. Dietrich, F.~Sannino, Phys. Rev. \textbf{D75}, 085018 (2007),
  \texttt{hep-ph/0611341}

\bibitem{Ohki:2010sr}
H.~Ohki, T.~Aoyama, E.~Itou, M.~Kurachi, C.J.D. Lin, H.~Matsufuru, T.~Onogi,
  E.~Shintani, T.~Yamazaki, PoS \textbf{LATTICE2010}, 066 (2010),
  \texttt{1011.0373}

\bibitem{Bursa:2010xn}
F.~Bursa, L.~Del~Debbio, L.~Keegan, C.~Pica, T.~Pickup, Phys. Lett.
  \textbf{B696}, 374 (2011), \texttt{1007.3067}

\bibitem{Hayakawa:2013maa}
M.~Hayakawa, K.I. Ishikawa, S.~Takeda, M.~Tomii, N.~Yamada, Phys. Rev.
  \textbf{D88}, 094506 (2013), \texttt{1307.6696}

\bibitem{Appelquist:2013pqa}
T.~Appelquist, R.~Brower, M.~Buchoff, M.~Cheng, G.~Fleming, J.~Kiskis, M.~Lin,
  E.~Neil, J.~Osborn, C.~Rebbi et~al., Phys. Rev. Lett. \textbf{112}, 111601
  (2014), \texttt{1311.4889}

\bibitem{Leino:2017lpc}
V.~Leino, J.~Rantaharju, T.~Rantalaiho, K.~Rummukainen, J.M. Suorsa,
  K.~Tuominen, Phys. Rev. D \textbf{95}, 114516 (2017), \texttt{1701.04666}

\bibitem{Leino:2017hgm}
V.~Leino, K.~Rummukainen, J.M. Suorsa, K.~Tuominen, S.~Tähtinen (2017),
  \texttt{1707.04722}

\bibitem{Capitani:2006ni}
S.~Capitani, S.~Durr, C.~Hoelbling, JHEP \textbf{11}, 028 (2006),
  \texttt{hep-lat/0607006}

\bibitem{Luscher:1991wu}
M.~Luscher, P.~Weisz, U.~Wolff, Nucl. Phys. \textbf{B359}, 221 (1991)

\bibitem{Luscher:1996vw}
M.~Luscher, P.~Weisz, Nucl. Phys. \textbf{B479}, 429 (1996),
  \texttt{hep-lat/9606016}

\bibitem{Omelyan:2003:SAI}
I.P. Omelyan, I.M. Mryglod, R.~Folk, {Computer Physics Communications}
  \textbf{151}, 272 (2003)

\bibitem{Rantaharju:2015yva}
J.~Rantaharju, T.~Rantalaiho, K.~Rummukainen, K.~Tuominen, Phys. Rev.
  \textbf{D93}, 094509 (2016), \texttt{1510.03335}

\bibitem{Luscher:2009eq}
M.~Luscher, Commun. Math. Phys. \textbf{293}, 899 (2010), \texttt{0907.5491}

\bibitem{Luscher:2010iy}
M.~Lüscher, JHEP \textbf{08}, 071 (2010), [Erratum: JHEP03,092(2014)],
  \texttt{1006.4518}

\bibitem{Cheng:2014jba}
A.~Cheng, A.~Hasenfratz, Y.~Liu, G.~Petropoulos, D.~Schaich, JHEP \textbf{05},
  137 (2014), \texttt{1404.0984}

\bibitem{Fritzsch:2013je}
P.~Fritzsch, A.~Ramos, JHEP \textbf{10}, 008 (2013), \texttt{1301.4388}

\bibitem{Fodor:2012td}
Z.~Fodor, K.~Holland, J.~Kuti, D.~Nogradi, C.H. Wong, JHEP \textbf{11}, 007
  (2012), \texttt{1208.1051}

\bibitem{Appelquist:2009ty}
T.~Appelquist, G.T. Fleming, E.T. Neil, Phys. Rev. \textbf{D79}, 076010 (2009),
  \texttt{0901.3766}

\bibitem{DeGrand:2009mt}
T.~DeGrand, A.~Hasenfratz, Phys. Rev. \textbf{D80}, 034506 (2009),
  \texttt{0906.1976}

\bibitem{Lin:2015zpa}
C.J.D. Lin, K.~Ogawa, A.~Ramos, JHEP \textbf{12}, 103 (2015),
  \texttt{1510.05755}

\bibitem{Hasenfratz:2016dou}
A.~Hasenfratz, D.~Schaich (2016), \texttt{1610.10004}

\bibitem{Capitani:1998mq}
S.~Capitani, M.~Luscher, R.~Sommer, H.~Wittig (ALPHA), Nucl. Phys.
  \textbf{B544}, 669 (1999), \texttt{hep-lat/9810063}

\bibitem{DellaMorte:2005kg}
M.~Della~Morte, R.~Hoffmann, F.~Knechtli, J.~Rolf, R.~Sommer, I.~Wetzorke,
  U.~Wolff (ALPHA), Nucl. Phys. \textbf{B729}, 117 (2005),
  \texttt{hep-lat/0507035}

\bibitem{Giusti:2008vb}
L.~Giusti, M.~Luscher, JHEP \textbf{03}, 013 (2009), \texttt{0812.3638}

\bibitem{Patella:2011jr}
A.~Patella, Phys. Rev. \textbf{D84}, 125033 (2011), \texttt{1106.3494}

\bibitem{Ryttov:2017toz}
T.A. Ryttov, R.~Shrock (2017), \texttt{1701.06083}

\bibitem{Amato:2015dqp}
A.~Amato, T.~Rantalaiho, K.~Rummukainen, K.~Tuominen, S.~Tähtinen, PoS
  \textbf{LATTICE2015}, 225 (2016), \texttt{1511.04947}

\end{thebibliography}

%%%%%%%%%%%%%%%%%%%%%%%%%%%%%%%%%%%%%%%%%%%%%%%%%%%%%%%%%%%%%%%%%%%%%%%%%%%%%
\end{document}